\documentclass[a4paper]{jpconf}
\usepackage[normalem]{ulem}
\usepackage{amsmath,amsfonts}
\usepackage{mathrsfs} \usepackage{graphicx,color} 

 \let\b=\beta \let\g=\gamma 
   
    \let\p=\pi
\let\s=\sigma  \let\f=\varphi \let\ph=\varphi
   
 \let\Th=\Theta  
   
\let\ee=\epsilon  \let\th=\theta \let\io=\infty

 \def\VV{{\cal V}}
\def\FF{{\cal F}} \def\HH{{\cal H}}

\def\GG{{\cal G}} \def\SS{{\cal S}}

\def\to{\rightarrow} \def\la{\left\langle} \def\ra{\right\rangle}

\newcommand{\beq}{\begin{equation}} \newcommand{\eeq}{\end{equation}}
\newcommand{\wh}{\widehat}

\begin{document}

\title{Packing hard spheres with short-range attraction in infinite
  dimension: \\ Phase structure and algorithmic implications}

\author{M Sellitto}

\address{DIII,
Seconda Universit\`a di Napoli, Real Casa dell'Annunziata, I-81031
Aversa (CE), Italy}

\author{F Zamponi} 

\address{LPT, \'Ecole Normale Sup\'erieure, UMR 8549 CNRS, 24 Rue
  Lhomond, 75005 Paris, France}

\begin{abstract}
We study, via the replica method of disordered systems, the packing
problem of hard-spheres with a square-well attractive potential when
the space dimensionality, $d$, becomes infinitely large.  The phase
diagram of the system exhibits reentrancy of the liquid-glass
transition line, two distinct glass states and a glass-to-glass
transition, much similar to what has been previously obtained by
Mode-Coupling Theory, numerical simulations and experiments.  The
presence of the phase reentrance implies that for a suitable choice of
the intensity and attraction range, high-density sphere packings more
compact than the one corresponding to pure hard-spheres can be
constructed in polynomial time in the number of particles (at fixed,
large $d$) for packing fractions $\f \leq 6.5 \, d \, 2^{-d}$.
Although our derivation is not a formal mathematical proof, we 
believe it meets the standards of rigor of theoretical physics,
and at this level of rigor it provides a small improvement of the lower bound on the
sphere packing problem. 
\end{abstract}


\section{Introduction}

Packing problems~\cite{TS10} are ubiquitous in science
  and engineering and arise in a variety of contexts ranging from
  biology (e.g. in the crowded cellular environment where
  physiological processes conspire significantly with excluded-volume
  effects~\cite{crowding}) to communication technology (e.g. in connection
  with the design of error correcting codes for signal transmission
  over noisy channels ~\cite{Hamming}).  From a
  mathematic~\cite{ConwaySloane,Co10} and
  algorithmic~\cite{TS10,TJ10,AS12,MT13} point of view the problem of
  packing sphere efficiently is most challenging when the dimension of
  the physical space $d$ becomes increasingly large.  This is mainly due
  to several unusual features which are at the origin of the so-called
  ``curse of dimensionality''.  It is well known, for example, that
  hypersphere packings cannot be very dense because the volume of the
  empty spaces left unoccupied by the spheres tends to become dominant
  in high-dimensional spaces thus giving a vanishingly small sphere
  packing density. 
 Lattice packings with simple unit cells and simple symmetries
  are particularly inefficient in filling space when $d$ is large~\cite{ConwaySloane}, hence it is reasonable
  to guess that dense packing with complex unit cells might be the densest ones in large $d$. However,
  such packings have complex symmetries and they are hard to construct explicitly in generic dimensions~\cite{ConwaySloane}.
  At the same time, numerical simulations require a minimal number of particles that grow exponentially with $d$~\cite{TJ10,AS12,MT13,CIPZ11}
  and become extremely hard for large $d$.
  In this situation it is by no means obvious whether
  the dense packings should have a periodic or rather an irregular
  structure and how to devise an algorithmic procedure to construct
  explicitly them~\cite{AS12}.
   
   In statistical physics, hard-sphere
systems have been used for a long time to describe the gas-liquid
transition (e.g. by van der Waals~\cite{VdW73}) and the geometric
structure of dense liquids (starting from Bernal~\cite{BM60}). More
recently, they have been much studied as models of colloids (with
steric and electrostatic stabilization).  The interest has been
especially motivated by the possibility of introducing a
(depletion-induced) attraction between colloidal particles by adding a
suitable amount of non-adsorbing polymers to the colloidal
solution. The intensity and the range of the attraction can be tuned
by changing the polymer concentration and the polymer coil radius,
respectively. By doing so, one can thus explore a wide range of static
and dynamic behaviors which are not directly accessible in a simple
liquid system. Some of the fascinating properties that have been
identified in these systems include the reversible melting-by-cooling
of the colloidal glass state and structurally distinct types of
dynamically arrested states~\cite{Sc02}.

The above
  colloid-polymer solutions can be modeled by an assembly of particles
  interacting through a potential made by a repulsive hard-core plus a
  very short-range attractive part.  In the following, we shall be
  focused on the phase behavior of such systems when the space
  dimensionality becomes infinitely large.  This is interesting for
several reasons.  First, in this limit the approach we use is arguably
exact~\cite{KPZ12} and this allows for a comparison with the results
obtained by alternative methods.
Moreover, our study provides evidence for the possibility of improving
slightly the lower bound on the sphere packing problem in large
dimensions.  

A short account of our work was presented in~\cite{noiEPL}.
In the present contribution we report extensively on the method we
have used and discuss thoroughly the results of our calculations.
In the remaining part of this introduction we explain further the
motivations behind our work, including their possible algorithmic
relevance. In section~\ref{sec:II}, we briefly outline the
Franz-Parisi effective potential method~\cite{FP98} (the related
replica calculation are reported in~\ref{app:A}).  In
section~\ref{sec:III} we present the derivation of the phase diagram
in an exemplary case and in the sticky limit as well.  We then
conclude by a qualitative description of the dynamical behavior
implied by our results, and by discussing further perspectives of our
work.

\subsection{Short-range attractive colloids: 
a re-entrant liquid-glass transition and two glasses}

Colloidal systems with short-range attraction display a very rich
phase diagram.  When the range of the attraction is not too short, the
phase diagram is characterized by a re-entrant liquid-glass transition
line. This means that for a suitable intensity of the attraction (or,
equivalently, temperature), the liquid-glass transition occurs at
density {\it higher} than for the pure hard-sphere system without
attraction.  Moreover, when the range of the attraction is short
enough, a glass-glass transition appears at high packing density. This
transition separates a ``repulsive glass'' phase dominated by the hard
core repulsion from an ``attractive glass'' phase dominated by the
short range attraction~\cite{Sc02}.  The attractive glass is
particularly interesting because it forms also at quite small
densities where the slowing down is induced by a ``gelation
mechanism'': filamentary chains of bonded particles form and percolate
the system.

This interesting phase diagram has been first obtained by
MCT~\cite{FP99,Fa99}, and later confirmed by numerical
simulations~\cite{ZP09,FSZT04,Da00}, experiments~\cite{EB02,Ph02},
and within the heterogeneous facilitation picture~\cite{Se10,Se12}.
MCT is often thought to be part of a more general Random First Order
Transition (RFOT) scenario for the glass transition. This idea is
based on the analogy between the glass problem and a class of spin
glass models~\cite{KT87,KW87b,WL12}: indeed, several studies found
glass-glass transitions in these spin glass (or lattice glass)
models~\cite{CCN04,CL06,Kr07,CL07,KTZ08} and gave further insight on
the relaxation dynamics around this transition~\cite{CCN04}.
According to the general RFOT scenario~\cite{WL12}, the same phase
diagram should be obtained using the replica method~\cite{MP99,WL12}.
The advantage of using the replica method is that one can also access
the glass phase (e.g. compute the equation of state of the
glass). Moreover, critical properties are easier to compute in the
replica formalism~\cite{FPRR11}.

So far only one attempt to use the replica method for this problem has
been reported~\cite{VPR06}.  It is based on the replicated Hypernetted
Chain (HNC) formalism, which is known to give a correct qualitative
phase diagram, even if its quantitative accuracy is rather poor.
Within this formalism the re-entrant glass transition was found,
however the glass-glass transition and in particular the attractive
glass phase were not found.  This is probably due to the fact that the
replicated HNC scheme is known to fail badly when the ``cage size'' is
very small~\cite{PZ10}, which is the case in the attractive
glass~\cite{Sc02}.  It is therefore important to carry out this
computation by using the ``small cage expansion''
scheme~\cite{MP99,PZ10}, which is more appropriate for this situation
and normally gives much better results from the quantitative point of
view.  In this paper we report such a computation and we show that its
results are consistent with the one obtained from MCT, which is nicely
consistent with the general RFOT picture.

\subsection{Large space dimension: 
algorithmic implications for the sphere packing problem}

In the following, we will stick to the $d\to\io$ limit, for two
reasons. First of all, computations are simplified, the theory can be
shown to be exact at the level of rigor of theoretical
physics~\cite{KPZ12}, and one is able to access all the transitions
that characterize the RFOT scenario (the dynamical transition, the
Kauzmann transition, and the jamming/glass close packing
point)~\cite{PZ10}.  A second and more important motivation is that
the presence of a re-entrant glass transition has an algorithmic
interest in $d$$\to$$\io$, which we now explain.

\paragraph*{Rigorous results on high dimensional sphere packings--}

The problem of packing spheres is very hard when
the space dimensionality becomes
larger than $d \sim 20$~\cite{ConwaySloane,Co10}. From an analytic
point of view, the recursive strategy
  of constructing optimal packings by slicing
$d$-dimensional packings in $d+1$ dimensions turns out to be very
inefficient~\cite{ConwaySloane,Co10} and the few good packings
which are known for dimension smaller than $d\sim 200$
have been mostly ``handcrafted''. Moreover, because in some
dimensions there are special symmetries that lead to the existence of
extremely dense packings, it is very hard to extrapolate the trend of
the maximum packing density to larger dimensions.  From the
algorithmic point of view, one could try to solve the problem by
adapting the ``simulated annealing'' procedure, which in this case
consists in constructing good packings by slowly compressing low
density configurations of spheres.  This procedure is known as the
Lubachevsky-Stillinger (LS) algorithm~\cite{LS90,SDST06}.
Unfortunately, as soon as $d>5$, crystallization becomes exceedingly
rare~\cite{SDST06,VCFC09,CIPZ11}: the system remains stuck ``forever''
in an amorphous phase, which at high enough density is a glass. It
seems that the time needed to crystallize
increases at least exponentially with dimension. Other smart
algorithms have been used to construct dense lattice
packings~\cite{TJ10,AS12,MT13} but unfortunately their running time
increases fast with dimension and for the moment the use of these
algorithms is limited to dimensions smaller than $d\sim 40$. 
It is also
important to stress that, at least to investigate amorphous packings,
the minimal number of particles that need to be studied increases exponentially
with $d$~\cite{CIPZ11}, which is an important limitation to study large
spatial dimensions.

Because finding optimal packings is so difficult for large $d$, only
{\it non-constructive} upper and lower bounds to the best packing
fraction (the fraction of volume covered by the spheres) $\f$ are
known. Roughly speaking (see~\cite{ConwaySloane,Co10,TS10} for more
precise formulations), the known best lower bound is $\f \geq (6d/e)
2^{-d}$~\cite{Va11} (note that it took 20 years to gain a factor of
$3/e$ with respect to the previous best lower bound $\f \geq 2d
2^{-d}$~\cite{Ba92}), while the known best upper bound is $\f \leq
2^{-0.5990 d}$~\cite{KL78}. Therefore, the upper and lower bounds are
exponentially divergent with $d$, leaving a huge gap open where the
densest packing might be located.

\paragraph*{Non-rigorous results on disordered packings--}
It has been proposed, based on an analysis of consistency conditions
for the pair correlation functions, that sphere packings could exist
up to a density $\f \sim 2^{-0.77865 d}$, i.e. somewhere in between
the lower and upper bounds but exponentially larger than the best
lower bound~\cite{TS10}. The fact that the pair correlation function
used in the analysis has no structure suggests that these packings, if
they exist, could be lattice packings with a very complex fundamental
cell or even disordered packings. Unfortunately, no way to construct
such packings, or to transform the conjecture in a more rigorous
bound, has been obtained.

The study of disordered packings in large space dimension is therefore
extremely interesting to obtain further insight into the problem.  The
analytical study of a system of amorphous hard spheres in large dimensions
within the RFOT scenario and using the replica
method~\cite{PZ06a,PZ10} gives the following predictions.

\begin{itemize}

\item
An ergodic liquid phase exists up to a density $\f_d$ that scales
asymptotically as $\f_d = 4.8 \, d \, 2^{-d}$.  At $\f_d$, the liquid
phase fragments into a large number of non-ergodic components.  Under
very general conditions, one can prove rigorously~\cite{MS06} that if
such an ergodicity breaking transition exists, the equilibration time
of any local dynamics is polynomial in the number of particles $N$ for
$\f < \f_d$, while it is exponential in $N$ for $\f > \f_d$.  Hence at
$\f_d$, a dynamical arrest of the MCT type towards a glass phase
happens.  Note however that this statement is not rigorous for hard
spheres because the very existence of $\f_d$ is not rigorously proven
in that case.

\item 
An exponential number of {\it amorphous} packings exist up to a
density $\f_{\rm GCP}$ scaling asymptotically as $\f_{\rm GCP} = d \,
\log(d) \, 2^{-d}$.  Yet, based on the statement above, the time
needed to explore these packings using standard local dynamics is
exponentially large in the number of particles.

\end{itemize}

This approach can be shown to be exact at the level of rigor of
theoretical physics~\cite{KPZ12}, and these predictions are consistent
with numerical simulations in dimensions $d$ ranging from $3$ to
$13$~\cite{CIPZ11,CCPZ12}.  We are therefore led to conclude that {\it
  exponentially many disordered sphere packings exist up to $\f_{\rm
    GCP} = d \, \log(d) \, 2^{-d}$, but sampling them using local hard
  sphere dynamics is exponentially hard in the number of particles $N$
  when the density is larger than $\f_d = 4.8 \, d \, 2^{-d}$}.  Note
that this situation resembles closely the one that is encountered in
many random optimization problems for which the RFOT scenario is
exact, for example the coloring of random graphs in the limit of a
large number of colors~\cite{ZK07}.  Note also that sampling the
packings uniformly is exponentially hard in $N$ for $\f > \f_d$, but
one can still construct packings in polynomial time in $N$ above $\f_d$ by
adiabatic compression of the packings at $\f_d$~\cite{CIPZ11} (again,
in close analogy with the coloring problem~\cite{KK07}).  This allows
to construct packings with density slightly larger than
$\f_d$, but probably still proportional to $d \, 2^{-d}$ (the
computation of the proportionality constant remains however an open
problem, and should be done by extending the replica method
following~\cite{KZ10}).

An interesting conclusion of the above discussion is that at the level
of rigor of theoretical physics, we can state that {\it disordered
  sphere packings exist and can be sampled with simulated annealing in
  polynomial time in $N$ at fixed $d$ for density at least equal to $\f_d = 4.8 \,
  d \, 2^{-d}$}, which provides a non-rigorous (but constructive)
small improvement of the lower bound of Vance $\f \geq (6d/e)
2^{-d}$~\cite{Va11} mentioned above 
(the reader should however keep in mind
that, unfortunately, the minimal number of particles $N$ must scale exponentially with $d$~\cite{CIPZ11},
which in practice renders the numerical simulations prohibitive for current computers above $d\sim 13$).  
In this paper we will show that
when a short range attraction is added to the hard sphere potential,
the glass transition line $\f_d$ moves to even higher densities.  The
presence of this re-entrance of the line $\f_d$ implies that one is
able to equilibrate the system in polynomial time in $N$ up to a
higher threshold. We will show that unfortunately the improvement is
only in the prefactor, which is brought to at most $\f_{d,max} \sim
6.5 \, d \, 2^{-d}$ for a suitable choice of the range and intensity
of the attraction.

\section{The Franz-Parisi potential and the replica method}
\label{sec:II}

We assume that the potential has a soft repulsive core of diameter
$D=1$.  The most general potential we want to consider in this paper
has the form 
\beq\label{eq:potential} v(r) = \ee (1-r)^2 \th(r<1) -
U_0 \th(1<r<1+\s) 
\eeq 
However, in the following we restrict for simplicity
to the case $\ee \to\io$ which corresponds to hard spheres with a
square-well attraction.
We scale the attraction width in such a way that $\wh\s = \s \, d$ is finite.
Moreover, the natural scale of packing fraction is the scaled
$\wh\f = 2^d \f/d$ where $\f$ is the packing fraction of the
repulsive core.

The basic idea of the replica approach to the glass transition is that
the kinetic slowing down on approaching the glass phase is due to the
sudden appearance of a bunch of long-lived metastable
states~\cite{KW87,KT88,KT89}. Under this assumption, the glass
transition can be detected by looking at the free energy of a system
constrained to be at a fixed distance from a reference equilibrium
configuration.  This is known as the {\it Franz-Parisi
  potential}~\cite{FP95,FP97,CFP98}, see also~\cite{KT89,Mo95} for an
alternative but very related approach.  In the present context, the
best way of making this construction explicit is the following. One
considers an equilibrium configuration $X = \{ x_i \}_{i = 1\cdots N}$
of the liquid at a given state point, and a second configuration $Y$
that is constrained to be closed to the first one, in such a way that
the mean square displacement \beq\label{eq:longMSD} \frac{d}N \la (
x_i - y_i)^2 \ra \leq 2 \wh A \ , \eeq where $\wh A$ is a fixed
constant. One then computes the free energy of the system $Y$ for
fixed $X$, and then averages it over the equilibrium distribution of
$X$. The result is the average free energy $V_{\rm FP}(\wh A)$ of a
system constrained to be at distance $\wh A$ from a typical liquid
configuration.  In the liquid phase, it is always possible for
particles to diffuse away from any reference configuration, and
correspondingly $V_{\rm FP}(\wh A)$ has a unique minimum at $\wh A =
\io$. On the contrary, in the glass phase, there is a metastable phase
-- corresponding to a local minimum of $V_{\rm FP}(\wh A)$ at finite
$\wh A$ -- in which the second system remains spontaneously close to
the first one, signaling the presence of a caging effect. The
secondary minimum appears discontinuously at a finite $\wh A$ below
the {\it dynamical transition} line $T_d(\f)$.  At the mean-field
level, this secondary minimum has an infinite life time.  Very
remarkably, such a construction can be done explicitly for realistic
models of glass formers, both analytically (using standard liquid
theory approximations) or
numerically~\cite{FP98,CFP98,DawFraSel03,CCGGGPV10}. From an analytic
point of view, the computation of the Franz-Parisi potential requires
the use of the replica method. This procedure has been developed
in~\cite{MP96,CFP98,MP09,PZ10}. For the specific case of the potential
in Eq.~\eqref{eq:potential}, the best known approximation has been discussed
(for $U_0=0$) in~\cite{BJZ11}. Moreover, it has been shown (at a
theoretical physics level, i.e. not rigorously) in~\cite{KPZ12} that
this approximation becomes exact in the limit $d\to\io$ where space
dimensionality becomes very large.  The extension of these results to
$U_0 \neq 0$ is straightforward and it is discussed in
\ref{app:A}.

In summary, we show in \ref{app:A} that 
\beq
V_{\rm FP}(\wh A) = \text{const.} + \frac{d}2 \, \VV_{\rm FP}(\wh A) \ ,
\eeq
and $\VV_{\rm FP}$ can be computed by means of
Eq.~\eqref{eq:VFPdio}. Moreover,
\beq \label{VVder}
\VV'_{\rm FP}(\wh A) = - \frac1{\wh
  A} \left[ 1 - \wh\f \FF_1(\wh A) \right] \ .  
  \eeq 
Therefore
 the stationary points (maxima and minima) of the Franz-Parisi potential
 are
located at values of $\wh A$ which are the solutions of
\beq
\label{eq:cage} 
\frac1{\wh\f} = \FF_1(\wh A) \ ,
\eeq 
with the
function $\FF_1(\wh A)$ defined in Eq.~\eqref{eq:F1phi}. 
The values of
$\wh A$ corresponding to the local minima of $V_{\rm FP}(\wh A)$
correspond to the long-time mean square displacement (i.e. the
Debye-Waller factor) in the glass phase.  In this context the
dynamical transition corresponds to the disappearance of all the local
minima of $V_{\rm FP}(\wh A)$, which happens when $\wh\f^{-1} >
\max_{\wh A} \FF_1(\wh A)$. Hence, the equation for the dynamical
transition is 
\beq
\label{phid} 
\frac1{\wh\f_d} = \max_{\wh A}
\FF_1(\wh A) \ .  
\eeq

\section{The static phase diagram of the square-well potential}
\label{sec:III}

Let us now discuss the {\it static} phase diagram, which is derived by
studying the behavior of the functions $\VV_{\rm FP}$ and $\FF_1$. The
connection with dynamics will be briefly discussed in the conclusions.

\subsection{Medium-range attraction: a single glass phase}

\begin{figure} 
\centering
\input{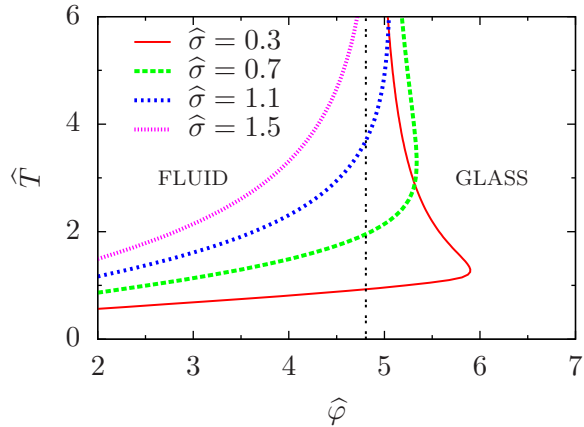}
\caption{Phase diagram in the rescaled variables, temperature
  $\wh{T}=k_{\scriptscriptstyle \rm B}T/U_0$ vs packing fraction
  $\wh\ph$, for moderately large square-well attraction width $\wh\s \geq 0.3$. The vertical
  line is the packing fraction at the dynamic glass transition for the
  purely hard-sphere potential. In this case there is only one glass
  phase but the fluid-glass transition line is reentrant: in a range
  of the control parameters the amorphous solid melts upon cooling
  (and the liquid freezes upon heating).
}
\label{phasediagram_sw}
\end{figure}

When the attraction width is large enough (above $\wh\s \simeq 0.19$),
the function ${\cal F}_1$ has a single maximum for all densities and
temperatures, and the resulting phase structure is easily determined:
for each temperature, a single glass phase exists for densities larger
than $\wh\f_d$ defined by Eq.~\eqref{phid}.  In
Fig.~\ref{phasediagram_sw} we show the phase diagram in rescaled
variables: reduced temperature, $\wh{T}=k_{\scriptscriptstyle \rm
  B}T/U_0$, and reduced packing fraction, $\wh\ph$. Interestingly,
there is a range of $\wh\ph$ above $\approx 5$ in which the glass
melts upon cooling and the resulting fluid freezes when temperature is
further lowered. This reentrance effect is driven by the width of the
square-well potential. The smaller the attraction width the deeper the
fluid phase enters into the glass region. This effect was already
found in~\cite{VPR06} by using the replica method in a different
approximation scheme.

\subsection{Short-range attraction: two glass phases}

When the attraction width is below $\wh\s \simeq 0.19$, the function
${\cal F}_1$ can have two maxima in some range of temperatures and densities,
and the determination of
the phase diagram requires some care in this case because of the
appearance of multiple glass phases.
To discuss in detail the several interesting features we find for the
square-well potential we focus on a representative case with $\wh\s
= 0.06$.

\begin{figure}[t]
\input{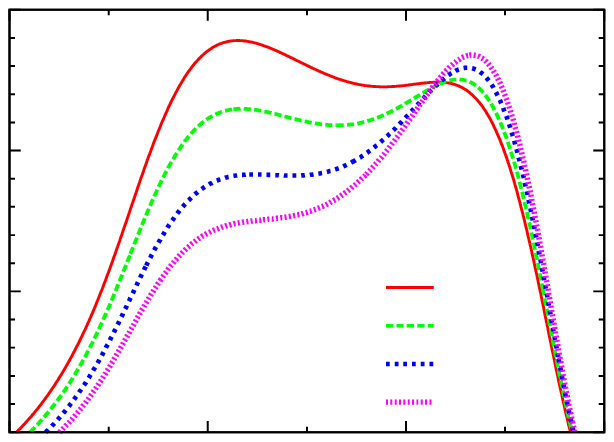}
\input{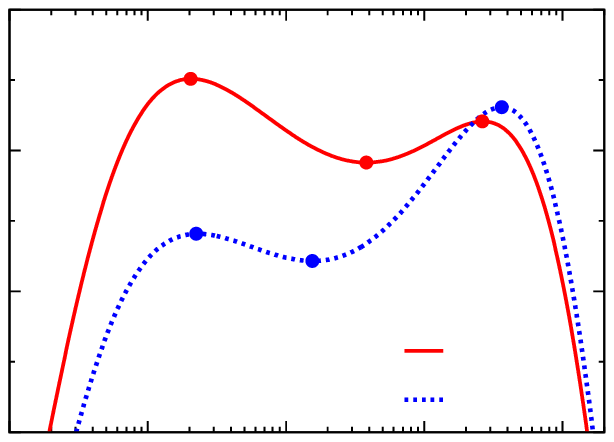}
\caption{ (Left) Shape of the function ${\cal F}_1$ for $\wh\s=0.06$ at some
  reduced temperatures $\wh{T}= k_{\scriptscriptstyle \rm B} T/U_0$. 
  (Right) The definitions
   of Eq.~\eqref{eq:spdef} are illustrated at temperatures $\wh{T}=0.5$ and $\wh T=0.47$.  
   In the
appropriate range of density $\max\{\wh\f^{\rm r}_{\rm d},\wh\f^{\rm
  a}_{\rm d} \} < \wh\f <\wh\f_{\rm sp}$, Eq.~\eqref{eq:cage} has four
solutions, two of them corresponding to the glasses and the two others
to local maxima of $\VV_{\rm FP}$. 
 }
\label{fig:F1A}
\end{figure}

\begin{figure} 
\input{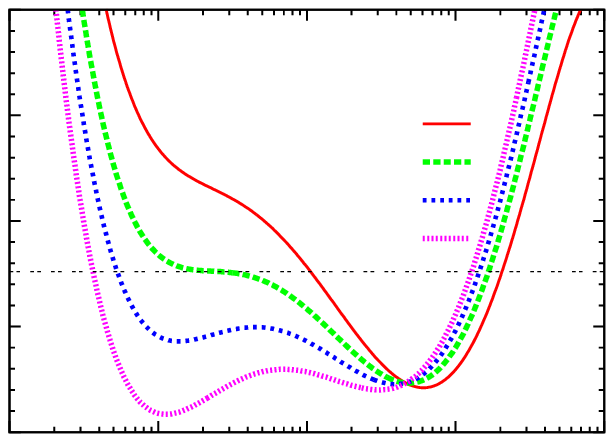}
\input{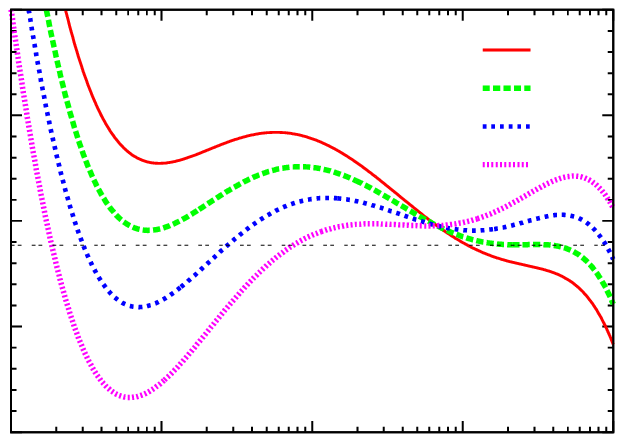}
\caption{
The Franz-Parisi potential $\VV_{\rm FP}(\wh A)$ for two temperatures,
$\wh{T}=0.5$ (left) and $\wh T=0.47$ (right), at several values of the density.
The corresponding functions $\FF_1$ are reported in Fig.~\ref{fig:F1A}.
}
\label{fig:VFP1}
\end{figure}

The function $\FF_1(\wh A)$ is reported in Fig.~\ref{fig:F1A} for
$\wh\s=0.06$ and two selected values of temperature. A crucial
observation is that $\FF_1$ does not depend on density.  It is seen
that in these cases, $\FF_1$ has two local
maxima separated by a local minimum.  Let us give the following
definitions, illustrated in Fig.~\ref{fig:F1A}: \beq\label{eq:spdef}
\FF_1 = \left\{
\begin{array}{ll}
1/\wh\f^{\rm a}_{\rm d} & \text{ at the maximum at smaller $\wh A$, } \\
1/\wh\f^{\rm r}_{\rm d} & \text{ at the maximum at larger $\wh A$, } \\
1/\wh\f_{\rm sp} & \text{ at the minimum; } \\
\end{array}
\right.  \eeq we also consider the analytic continuations of these
densities when they exist.  Therefore in some range of densities
Eq.~\eqref{eq:cage} has four solutions: two of them (the first and the
third, upon increasing $\wh A$) correspond to local minima and the two
others correspond to local maxima of the Franz-Parisi potential. This
is illustrated in Fig.~\ref{fig:VFP1}, and remember that while $\FF_1$
does not depend on density, $\VV_{\rm FP}$ does.  The two local minima
correspond to two different metastable glass states, and we call them
the {\it attractive glass} (the one with smaller $\wh A$) and {\it
  repulsive glass} (with larger $\wh A$). Therefore, in this region of
density we have coexistence between the two glasses.

\begin{figure} 
\centering
\input{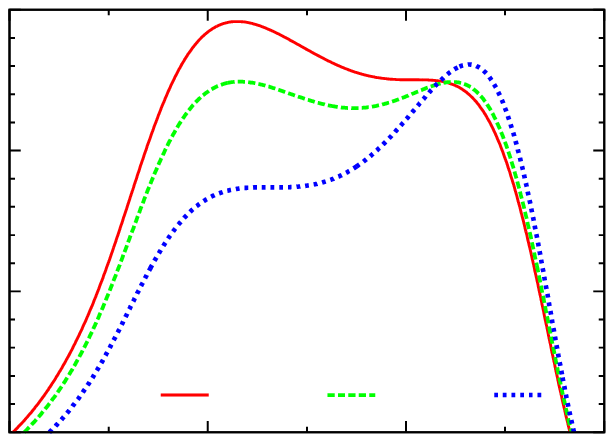}
\caption{ Shape of the function ${\cal F}_1$ for $\wh\s=0.06$ for the
  special temperatures $\wh{T}_{{\mathsf A}^{\rm r}_3}$,
  $\wh{T}_{\mathsf C} $, $\wh{T}_{{\mathsf A}^{\rm a}_3}$ defined in
  the text.  }
\label{fig:F1A_special}
\end{figure}

\begin{figure}[h] 
\begin{minipage}[b]{0.45\textwidth}
\input{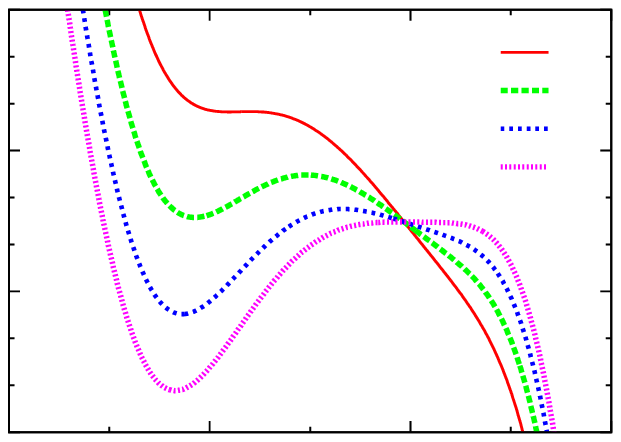}
\input{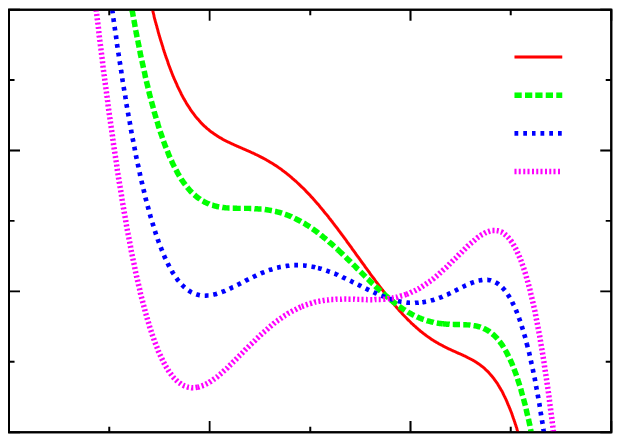}
\input{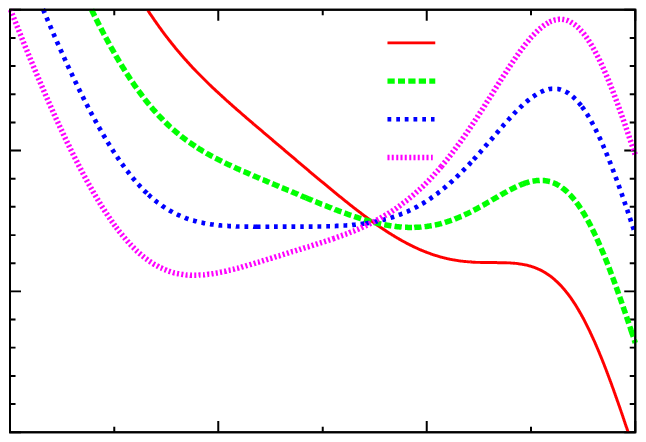}
\end{minipage}\hspace{2pc} 
\begin{minipage}[b]{0.45\textwidth}
\caption{ Shape of the Franz-Parisi potential $\VV_{\rm FP}$ for
  $\wh\s=0.06$ for the special temperatures $\wh{T}_{{\mathsf A}^{\rm
      r}_3}$ (top), $\wh{T}_{\mathsf C} $ (middle), $\wh{T}_{{\mathsf
      A}^{\rm a}_3}$ (bottom) defined in the text, for several
  densities at each temperature.  The corresponding functions $\FF_1$
  are reported in Fig.~\ref{fig:F1A_special}.  }
\label{fig:VFP_special}
\end{minipage}
\end{figure}

To obtain the phase diagram we must analyze the behavior of $\FF_1(\wh
A)$ systematically as a function of temperature.  From
Fig.~\ref{fig:F1A} we deduce the existence of several different
temperature ranges:
\begin{itemize}
\item For $\wh T < \wh{T}_{{\mathsf A}^{\rm r}_3}$, $\FF_1$ has a
  unique maximum at small $\wh A$, corresponding to $\wh\f^{\rm
    a}_{\rm d} $.  The other two densities in~\eqref{eq:spdef} do not
  exist, only the attractive glass exists for these temperatures, in
  the region $\wh\f \in [ \wh\f^{\rm a}_{\rm d},\io)$.
\item
For $ \wh T = \wh{T}_{{\mathsf A}^{\rm r}_3}$, a pair of stationary
points of $\FF_1$ (a local minimum and a local maximum at large $\wh
A$) appear (Fig.~\ref{fig:F1A_special}): this leads to $\wh\f_{\rm sp}
= \wh\f^{\rm r}_{\rm d}$.  At this temperature, the repulsive glass
solution exists only marginally: it corresponds to a very flat maximum
of the Franz-Parisi potential, see Fig.~\ref{fig:VFP_special}(top).
\item For $ \wh{T}_{{\mathsf A}^{\rm r}_3} < \wh T < \wh{T}_{\mathsf
  C}$, there are two local maxima separated by a local minimum, and
  the maximum at small $\wh A$ is higher. In other words, $\wh\f^{\rm
    a}_{\rm d} < \wh\f^{\rm r}_{\rm d}$.  In this region the
  attractive glass exists for $\wh\f \in [ \wh\f^{\rm a}_{\rm
      d},\io)$, while the repulsive glass exists for $\wh\f \in [
      \wh\f^{\rm r}_{\rm d} , \wh\f_{\rm sp} ]$.  The Franz-Parisi
    potential is reported in Fig.~\ref{fig:VFP1}(right).
\item At $\wh{T}_{\mathsf C}$ the two maxima of $\FF_1$ have the same
  height (Fig.~\ref{fig:F1A_special}), hence the two lines cross
  $\wh\f^{\rm a}_{\rm d} = \wh\f^{\rm r}_{\rm d}$. The Franz-Parisi
  potential is in Fig.~\ref{fig:VFP_special}(middle).
\item
For $\wh{T}_{\mathsf C} < \wh T < \wh{T}_{{\mathsf A}^{\rm a}_3}$ the
situation is reversed, $\wh\f^{\rm a}_{\rm d} > \wh\f^{\rm r}_{\rm
  d}$, but for the rest nothing changes.  The Franz-Parisi potential
is in Fig.~\ref{fig:VFP1}(left).
\item When $ \wh T = \wh{T}_{{\mathsf A}^{\rm a}_3}$, the local
  minimum and the small-$\wh A$ maximum of $\FF_1$ coalesce
  (Fig.~\ref{fig:F1A_special}).  Here $\wh\f_{\rm sp} = \wh\f^{\rm
    a}_{\rm d}$ and the attractive glass becomes a marginally stable
  point.  The Franz-Parisi potential is in
  Fig.~\ref{fig:VFP_special}(bottom).
\item
Finally, for $\wh T > \wh{T}_{{\mathsf A}^{\rm a}_3}$ there is a
unique maximum at large $\wh A$ corresponding to the attractive glass,
which is the only state in this region and exists for $\wh\f \in [
  \wh\f^{\rm r}_{\rm d},\io)$.
\end{itemize}

\begin{figure} 
\input{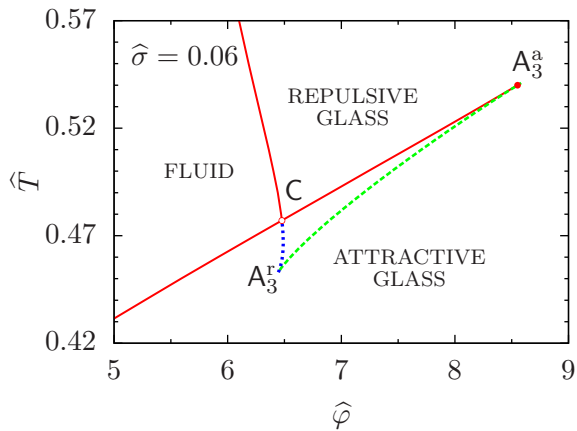}
\centering
\caption{Phase diagram for the representative case $\wh\s=0.06$.
  There are two distint types of glass phases and fluid-glass
  transitions plus a glass-glass transition line.  The two fluid-glass
  transition lines meet at a crossing point ${\mathsf C}$ along with
  the glass-glass transition line.  The line ${\mathsf C}$-${\mathsf
    A}^{\rm a}_3$ corresponds to a spinodal for the attractive glass,
  above which this phase does not exist anymore. Similarly, the lines
  ${\mathsf C}$-${\mathsf A}^{\rm r}_3$ and ${\mathsf A}^{\rm
    r}_3$-${\mathsf A}^{\rm a}_3$ are spinodals of the repulsive
  glass.  The glass-glass coexistence region is delimited by these
  three lines.  The two special points ${\mathsf A}^{\rm a}_3$ and
  ${\mathsf A}^{\rm r}_3$ are cusp singularities; they correspond to
  points where the two glasses coalesce giving rise to a higher order
  singularity in the Franz-Parisi potential.  }
\label{sigma006}
\end{figure}

In Fig.~\ref{sigma006} we show the transition lines in the phase
diagram that delimit the regions defined above.  The two lines
$\wh\f^{\rm a}_{\rm d}(\wh T)$ and $\wh\f^{\rm r}_{\rm d}(\wh T)$
cross each other at point ${\mathsf C}$.  Below $\wh{T}_{\mathsf C}$,
the liquid-glass transition is given by $\wh\f^{\rm a}_{\rm d}(\wh T)$
and the liquid transforms into the attractive glass, while above
$\wh{T}_{\mathsf C}$, the liquid-glass transition is given by
$\wh\f^{\rm r}_{\rm d}(\wh T)$ and the liquid transforms into the
repulsive glass.  Hence \beq \wh\f_{\rm d}(\wh T) = \min\{ \wh\f^{\rm
  r}_{\rm d}(\wh T), \wh\f^{\rm a}_{\rm d}(\wh T) \} \ .  \eeq and the
liquid is ergodic for $\wh\f < \wh\f_{\rm d}(\wh T)$.

For $\wh\f > \wh\f_{\rm d}(\wh T)$, the system is arrested but the two
different glass phases can exist. The line $\wh\f^{\rm a}_{\rm d}(\wh
T)$ can be continued from point ${\mathsf C}$ to point ${\mathsf
  A}^{\rm a}_3$ where it merges with the line $\wh\f_{\rm sp}$.  The
point ${\mathsf A}^{\rm a}_3$ is a critical endpoint representing a
cusp singularity that corresponds to a very singular point of the
Franz-Parisi potential.  Similarly, the line $\wh\f^{\rm r}_{\rm
  d}(\wh T)$ can be continued from point ${\mathsf C}$ to point
${\mathsf A}^{\rm r}_3$ where it merges with the line $\wh\f_{\rm
  sp}$.  Together, these three lines delimit the region of temperature
and density where the two glasses coexist, while outside this region
only one glass is present. This scenario is the standard one
associated with a first order phase transition between the two
glasses, as described by the Franz-Parisi potential that exhibits two
stable minima in the coexistence region.

\subsection{The sticky limit}

It is interesting to consider a sticky sphere limit where $\wh\s \to
0, \, U_0 \to \infty$ while $\mu = -\b U_0 - \log \wh\s$ is kept
constant.  In this limit the control parameters are therefore $\wh\f$
and $\mu$.  One can show (\ref{app:A}) that in this limit the
small-$\wh A$ maximum of $\FF_1(\wh A)$, corresponding to the
attractive glass, moves to $\wh A=0$. Because of the sticky
attraction, particles do not move at all in the attractive glass
phase. The height of this maximum can be computed explicitly and gives
$\wh\f^{\rm a}_{\rm d} = 2 e^{\mu}$, see Eq.~\eqref{phid0}.  Hence in
this case the spinodal point of the attractive glass can be computed
analytically and reaches arbitrarily large densities, therefore the
point ${\mathsf A}^{\rm a}_3$ moves to infinite density.  Still, the
other two lines $\wh\f^{\rm r}_{\rm d}$ and $\wh\f_{\rm sp}$
--corresponding respectively to the large-$\wh A$ local maximum and
the local minimum-- have to be computed numerically. The resulting
phase diagram is reported in Fig.~\ref{sticky}.

\subsection{The evolution of the phase diagram with $\wh\s$: 
${\mathsf A}_3$ and ${\mathsf A}_4$ singularities}

\begin{figure}[t] 
\centering
\input{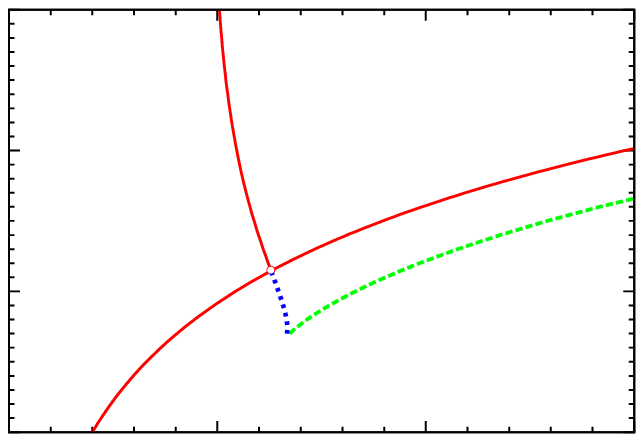}
\caption{Phase diagram in the sticky limit $\wh\s \to 0, \, U_0 \to
  \infty$ while keeping $\mu = -\b U_0 - \log \wh\s$ constant. }
\label{sticky}
\end{figure}

We now summarize the evolution of the phase diagram when the range of
the attraction $\wh\s$ is changed.  At large $\wh\s$, the phase
diagram displays a single re-entrant liquid-glass transition line, see
Fig.~\ref{phasediagram_sw}.  At very small $\wh\s$, a phase
coexistence region between two glasses is observed, and is delimited
by the triangle ${\mathsf A}^{\rm a}_3$-${\mathsf C}$-${\mathsf
  A}^{\rm r}_3$ in Fig.~\ref{sigma006}. This triangle is quite large
at very small $\wh\s$ (its size diverges in the sticky limit, see
Fig.~\ref{sticky}). Its size decreases with increasing $\wh\s$, until
it disappears at $\wh\s \approx 0.19$ where the three points ${\mathsf
  A}^{\rm a}_3$-${\mathsf C}$-${\mathsf A}^{\rm r}_3$ merge into a
further higher-order singularity of type ${\mathsf A}_4$, also known
as swallowtail bifurcation.

The higher-order singularities, ${\mathsf A}_3$ and ${\mathsf A}_4$,
are special singular points of the Franz-Parisi potential.  From
Eq.~\eqref{VVder} it is easy to show that if Eq.~\eqref{eq:cage} is
satisfied (i.e. on stationary points of the potential), the vanishing
of the $n$-th derivative of $ \FF_1(\wh A)$ leads to the vanishing of
the $(n+1)$-th derivative of $\VV_{\rm FP}(\wh A)$. On the points
${\mathsf A}_3$, the first two derivatives of $\FF_1$ vanish (see
Fig.~\ref{fig:F1A_special}) and therefore the first three derivatives
of the Franz-Parisi potential vanish (see Fig.~\ref{fig:VFP_special}).  On
the point ${\mathsf A}_4$, the first three derivatives of $\FF_1$
vanish and therefore the first four derivatives of the Franz-Parisi
potential vanish (not shown). This higher-order singularities lead to
peculiar properties of the relaxation dynamics, that have been
explored in great detail in the Mode-Coupling theory
framework~\cite{BF99,Fa99,Da00,CCN04,Go09} and numerically~\cite{STZ03,CR07}.
However, note that point ${\mathsf A}^{\rm r}_3$ is a {\it maximum} of the Franz-Parisi
potential and therefore it is unstable, see Fig.~\ref{fig:VFP_special}(top). This instability has also been discussed within Mode-Coupling
Theory~\cite{Go09}.

\subsection{The Kauzmann transition}

Finally, it is important to stress that the complexity contains a term
proportional to $\log(d)$, which makes it positive for all finite
values of $\wh\f$. Hence the Kauzmann transition happens at $\wh\f \to
\io$ and $\wh A\to 0$~\cite{PZ10} and it is therefore out of the range
of the phase diagrams discussed above. In particular, this implies
that the Kauzmann transition always happens in a region of the phase
diagram where there is no coexistence between different glasses.  The
complexity associated to both glasses is therefore positive and for this reason
we did not discuss the Kauzmann transition.

\section{Conclusion}

The main result of this paper is that a static replica picture based
on the Franz-Parisi potential allows to re-derive many of the results
that have been previously obtained using Mode-Coupling theory for
attractive colloids~\cite{Da00,Sc02,Go09}, namely the re-entrance of
the glass transition line and the coexistence of two glass phases for
very short range attractions. Here we limited ourselves to the
$d\to\io$ limit where computations are easier and the replica theory
is exact~\cite{KPZ12}, but the computations can in principle be
generalized to any finite dimension~\cite{PZ10}. Moreover, we found
that for suitable attraction range $\wh\s \approx 0.06$ and
temperature $\wh T \approx 0.48$, the fluid phase extends to density
$\wh\f \approx 6.5$. This shows that sphere packings exist up to
packing fraction $\f \approx 6.5 \, d \, 2^{-d}$ and can be produced
by slow compression of the fluid phase of attractive systems with the
above parameters.

Let us conclude by some speculations on the connection between the
static replica picture presented in this paper and the dynamic MCT
picture of~\cite{Da00,Sc02,Go09}, and by presenting some perspective
for future work.

\subsection{Dynamics}

It is very important to discuss the consequences that the static phase
diagram obtained above has for dynamics.  This is crucial to compare
theory with numerical simulations and experiments, that in the glass
phase are necessarily out of equilibrium. In simple spin glass models
this connection is perfectly understood (both at equilibrium and out
of equilibrium) within the general RFOT
framework~\cite{KW87,KT88,KT89,CK93,CC05,WL12}.  Unfortunately, when
one tries to translate this to approximate theories of particle
systems based on MCT and replicas, the situation is much more
complicated and controversial results have been obtained in the
past~\cite{KW87,IM10,SS10,Sz10,PZ10,CIPZ11}. This is due to the fact
that, although the general RFOT picture is perfectly consistent at
least for large enough dimensions~\cite{CIPZ12}, its practical
implementation using liquid theory approximate closure can lead to
inconsistencies (a fact which is well known in the theory of simple
liquids~\cite{Hansen}).

Moreover, the discussion of out of equilibrium dynamics is not obvious
because in general this dynamics depends a lot on the preparation
history and the protocol that is used to explore the glass phase. This
is particularly true in the present situation where two distinct glass
phases can coexist.  In principle, if one considers a slow annealing
in the glass phase, one can obtain precise information through the
so-called {\it state-following} method~\cite{KZ10}. However, such a
computation goes beyond the scope of this work.

Given these important remarks, we can discuss qualitatively the
evolution of the cage radius in the glass phase based on our results
for the Franz-Parisi potential.
When $\wh\f < \wh\f_{\rm d}(\wh T)$, the system is in a liquid phase
and it is ergodic. Particles can diffuse and the long time limit of
connected density correlations is zero. As usual, when $\wh\f \to
\wh\f_{\rm d}(\wh T)$ from below, dynamics slows down until at
$\wh\f_{\rm d}(\wh T)$ diffusion is arrested and the long time limit
of the mean square displacement becomes equal to a constant according
to Eq.~\eqref{eq:longMSD}. For densities very close to $\wh\f_{\rm
  d}(\wh T)$, the mean square displacement plotted as a function of
time exhibits a long plateau before crossing over to the diffusive
regime, and the value of this plateau corresponds roughly to the long
time mean square displacement in the glass phase.

We therefore have two different behavior depending on
temperature. When $\wh T < \wh{T}_{\mathsf C}$, the liquid arrests
into the attractive glass and the value of mean square displacement
corresponds to the small $\wh A$ solution. Instead, when $\wh T >
\wh{T}_{\mathsf C}$ the liquid arrests into the repulsive glass with
larger $\wh A$.

Suppose that we now continue to compress the system slowly and
isothermally, out of equilibrium into the glass phase.  We have
several possibilities:

\begin{itemize}

\item At low $\wh T < \wh{T}_{{\mathsf C}}$, the system jumps in the
  attractive glass at $\wh\f_{\rm d}$.  Then it remains in this phase
  upon compression. This is because, even if the repulsive glass
  solution can appear in some range of density, dynamics is arrested
  in the attractive glass, which does not have a spinodal that could
  make it unstable. Therefore the blue and green lines in
  Fig.~\ref{sigma006} cannot be observed under isothermal compression
  below $\wh{T}_{{\mathsf C}}$.
  
\item At intermediate temperature $\wh{T}_{{\mathsf C}} < \wh T <
  \wh{T}_{{\mathsf A}^{\rm a}_3}$ the system is stuck in the repulsive
  glass at $\wh\f_{\rm d}$.  However upon further increasing density,
  the repulsive glass solution disappears at large enough density
  (corresponding to $\wh\f = \wh\f_{\rm sp}(\wh T)$, the green line in
  Fig.~\ref{sigma006}).  Around this point, the system undergoes a
  transition to the attractive glass, which then remains stable upon
  further compression.  Note however that strong history dependent out
  of equilibrium effects are surely present, therefore the transition
  can happen everywhere between the red and green lines in
  Fig.~\ref{sigma006}.

\item At $\wh T > \wh{T}_{{\mathsf A}^{\rm a}_3}$, there is no
  attractive glass.  The system jumps into the repulsive glass at
  $\wh\f_d$ and remains there at any other density because this is the
  only solution.

\end{itemize}

In general, it is reasonable to expect that the repulsive glass phase
will be very hard to observe below $\wh{T}_{{\mathsf C}}$. Although
this phase exists in a strictly mean-field picture as a metastable
minimum, one can easily see from Fig.~\ref{fig:VFP1}(bottom) that the
barriers separating this phase from the liquid and the attractive
glass are very small. It is therefore most probable that in finite
dimensions these barriers are easily crossed in this region, leading
to a very short lifetime of the repulsive glass.  The most relevant
line in the triangle in Fig.~\ref{sigma006} seems therefore the line
${\mathsf A}^{\rm a}_3$-${\mathsf C}$, consistently with MCT
results~\cite{Da00,Go09}.

In particular, the ${\mathsf A}^{\rm r}_3$ singularity corresponds to
the point where the repulsive glass disappears leading to a {\it local
  quartic maximum} of the Franz-Parisi potential, as shown by the
curve for $\wh\f=6.44$ in Fig.~\ref{fig:VFP_special}(top): it is
therefore dynamically unobservable as the repulsive glass will always
be unstable around this singularity.  This is consistent with results
from MCT~\cite{Go09}.  On the contrary, the ${\mathsf A}^{\rm a}_3$
singularity corresponds to a {\it local quartic minimum} of the
Franz-Parisi potential, see the curve for $\wh\f=8.56$ in
Fig.~\ref{fig:VFP_special}(bottom). This leads to a unique glass phase
with peculiar dynamical properties and a very slow logarithmic
relaxation, as it has been shown within MCT~\cite{Go09}.

\subsection{Perspectives}

The discussion of dynamics reported above is very qualitative and
preliminary.  Exploring further the connection between static and
dynamical pictures will surely lead to a better understanding of the
rich out of equilibrium dynamics in the glass-glass coexistence phase,
where one could naturally expect hysteresis effects. We believe that
such a study is a very interesting subject for future work. Performing
numerical simulations in high dimensions, following~\cite{CIPZ11},
could be very useful to remove undesired metastability effects that
complicate the picture. Of course, the hope is to put back these
effects in a controlled way once the mean field picture is fully
understood.

Our work opens the way to several other studies.  First of all one
should understand better the nature of the glass phase close to the
glass-glass transition. It is to be expected that the complexity
function will have two distinct branches, corresponding to the two
different sets of glassy states in coexistence.  The solution that
maximizes the complexity should then correspond to the ``typical''
phase, in which equilibrium configurations will be found with
probability 1 in the thermodynamic limit. This corresponds as usual to
choosing the solution that minimizes the Franz-Parisi potential and
would provide an ``equilibrium'' definition of the glass-glass
transition line in the coexistence region. How activated barrier
processes change the picture within RFOT theory remains an open
problem that would be very interesting to investigate.  Another
question that would be important to address is the role of the
Kauzmann transition in finite dimension, which we did not discuss here
because in the limit $d\to \io$ this transition moves at infinite
$\wh\f$, and is therefore outside the range of relevant densities for
the glass-glass coexistence region.  Furthermore, one could compute
the equation of state of the two glasses, the jump of specific heat at
the glass transition, and so on.

For soft matter applications, one would like of course to repeat this
calculation in $d=3$. We expect that the qualitative phase diagram
will remain the same in low dimensions. Unfortunately, for the moment
the small cage expansion scheme does not give good quantitative
results for the dynamical transition in low dimensions, but there is
hope to improve it.

For algorithmic applications, it would be very important to perform a
``state following'' calculation~\cite{KZ10}, to study how much the
states at $\f_d$ can be compressed adiabatically: this would give the
true threshold beyond which packings cannot be constructed
in polynomial time by simulated annealing.

\paragraph*{Acknowledgments --}
We warmly thank P.~Charbonneau and E.~Zaccarelli for stimulating
discussions.

\appendix

\section{Gaussian replica equations for a generic potential}
\label{app:A}

We collect here all the equations that are needed to obtain the
results presented in this work.  Because these are based on extensions
of previous work~\cite{PZ10,BJZ11}, we do not provide the general
derivation but rather highlight the extensions that are needed with
respect to the previous works.

We assume that the potential has a soft repulsive core of diameter
$D=1$.  The most general potential we consider has the form 
\beq 
v(r) = \ee (1-r)^2 \th(r<1) - U_0 \th(1<r<1+\s) 
\eeq 
Here $T$ is temperature and $\b = 1/(k_B T)$.
We define $\wh\s
= \s \, d$ and $\wh\ee = \b\ee/d^2$ and $\wh\f = 2^d \f/d$ where $\f$ is
the packing fraction of the repulsive core.
We also define $\b U_0 = \wh U_0 = 1/\wh T$.

\subsection{Finite dimensions}

The approximation scheme used here holds for $\b\ee$ large enough and
is based on~\cite[Eq.~(22) and (23)]{BJZ11}, which give the replicated
free entropy separated between the harmonic and the liquid
contributions \beq\label{st1}
\begin{split}
\SS(m,A)&= S_h(m,A)+\SS_{\rm liq}(T/m,\f)  
+ 2^{d-1} \f y_{\rm liq}^{\rm HS}(\f) G_m(A) \ , \\
G_m(A) &= d \int_0^\io dr \, r^{d-1} \, [ q(A;r)^m - e^{-\b m v(r)}] \ , \\
S_h(m,A) &= \frac{d}2 (m-1) \log(2\pi A) - \frac{d}2 (1-m-\log m) \ ,
\end{split}
\eeq for $m$ replicas at temperature $T$, in a Gaussian cage of
variance $2A$.  The function $q(A;r) = \int d^dr' \g_{2A}(\vec r')
e^{-\b v(|\vec r - \vec r'|)}$ is defined in~\cite[just after
  Eq.~(16)]{BJZ11} where $\gamma_{2A}$ is a normalized and centered
Gaussian of variance $2A$, and $y_{\mathrm{liq}}^{\mathrm{HS}}$ is the contact value of the
hard-spheres cavity function.  Introducing bipolar coordinates, as
in~\cite[Appendix C.2.a]{PZ10}, we obtain the generalization
of~\cite[Eq.~(C16)]{PZ10} to a generic potential $v(r)$ \beq
\label{st2}
\begin{split}
q(A;r) &= \int_0^\io du \, e^{-\b v(u)} \left(\frac{u}r
\right)^{\frac{d-1}2} \frac{ e^{ -\frac{(r-u)^2}{4A} }}{\sqrt{4\p A}}
\left[ e^{ -\frac{ru}{2A} } \sqrt{\pi \frac{ru}{A}}
  I_{ \frac{d-2}{2} } \left( \frac{ru}{2A} \right)\right] \ .
\end{split}\eeq
From the replicated entropy we can obtain the Franz-Parisi potential
\beq\begin{split} 
\b V_{\rm FP}(A) &= - \left. \frac{d(\SS/m)}{dm}
\right|_{m=1} = S_{\rm liq}(T) -d - \frac{d}2 \log(2\pi A) -
2^{d-1} \f y_{\rm liq}^{\rm HS}(\f) H_1(A) \ , \\ H_m(A)& = m
\frac{\partial G_m(A)}{ \partial m} \ .
\end{split}\eeq
Note that the Franz-Parisi potential corresponds to the complexity at
$m=1$ as a function of $A$.

The local minima of $V_{\rm FP}(A)$ correspond to stable glass phases.
Because $\SS(m,A)$ is independent of $A$ at $m=1$, one can show that
the stationary points in $A$ of $V_{\rm FP}(A)$ are the same as those
of $\SS(m,A)$ for $m\to 1$.  In general, the stationary points of
$\SS(m,A)$ are given by the condition \beq \frac{d}{2^d \f y_{\rm
    liq}^{\rm HS}(\f)} = \frac{A}{1-m} \frac{\partial G_m(A)}{
  \partial A} \equiv F_m(A) \ .  \eeq

\subsection{Infinite dimension}

We follow \cite{PZ10}: in the limit of infinite dimension, we are
interested in the scaling $A = \wh A/d^2$.  One can show that 
\beq
\lim_{d\to \io} e^{-d^2 z} \sqrt{2 \pi d^2 z} I_{\frac{d-2}2}(d^2 z) =
e^{-\frac{1}{8 z}} \ .  
\eeq 
Using this result in (\ref{st2}), we have
\beq q(A;r) = \int_0^\io du \, e^{-\b v(u)} \left( \frac{u}{r}
\right)^{\frac{d-1}{2}} \frac{ e^{ -\frac{(r-u)^2}{4A} }}{\sqrt{4\p
    A}} e^{ -\frac{ \wh A}{4 r u} } 
\eeq 
Changing variables again to
$t = \frac{r-1}{\sqrt{4A}}$ and $s = \frac{u-1}{\sqrt{4A}}$ and using
\beq \left( \frac{u}{r} \right)^{\frac{d-1}{2}} = \left( \frac{1 +
  \frac{s \sqrt{4\wh A}}d }{ 1 + \frac{t \sqrt{4\wh A}}d }
\right)^{\frac{d-1}{2}} \sim e^{(s-t) \sqrt{ \wh A}} \ , 
\eeq 
we get
\beq \label{F0dinf}
\begin{split} q(\wh A;t) &\sim e^{-\frac{\wh A}{4}}
  \frac1{\sqrt{\pi}} \int_{-\io}^\io ds \, e^{-\b v\left(1 + s \sqrt{4
      \wh A}/d \right)} e^{(s-t) \sqrt{\wh A} - (t-s)^2} \\ & =
  e^{-\frac{\wh A}{4}} \frac1{\sqrt{\pi}} \int_{-\io}^0 ds \, e^{-4
    \wh A\wh\ee s^2 + (s-t) \sqrt{ \wh A} - (t-s)^2} +
  e^{-\frac{\wh A}{4}} \frac1{\sqrt{\pi}} \int_{0}^{\wh\s/\sqrt{4\wh
      A}} ds \, e^{\wh U_0 +(s-t) \sqrt{ \wh A} - (t-s)^2} \\ & +
  e^{-\frac{\wh A}{4}} \frac1{\sqrt{\pi}} \int_{\wh\s/\sqrt{4 \wh
      A}}^\io ds \, e^{(s-t) \sqrt{ \wh A} - (t-s)^2} \ .
\end{split}\eeq

We can use this result to compute $\GG_m(\wh A) = \lim_{d\to\io} G_m(A d^2)$:
\beq\label{taudinf}
\begin{split}
\GG_m(\wh A) &= \lim_{d\to\io} \sqrt{4 \wh A} \int_{-d/\sqrt{4\wh A}}^\io dt 
\left(1 + t\frac{ \sqrt{4\wh A}}{d} \right)^{d-1} 
 \left[   q(\wh A;t)^m - e^{-\b m v\left(1+t\sqrt{4\wh A}/d\right)} \right] \\ & 
= \int_{-\io}^\io dy \, e^y 
 \left[   q(\wh A;y)^m - e^{-\b m v\left(1+ y/d\right)} \right]
 \\
&=   \int_{-\io}^0 dy \, e^{y} 
\left[   q(\wh A;y)^m -
e^{ - m \wh\ee  y^2   }  \right] 
+  \int_{0}^{ \wh\s} dy \, e^{y} 
\left[   q(\wh A;y)^m -
e^{ m \wh U_0 }  \right] \\
&+   \int_{ \wh\s }^\io dy \, e^{y} 
\left[   q(\wh A;y)^m - 1 \right] 
\end{split}
\eeq
where $q(\wh A;y) = q(\wh A;t/\sqrt{4\wh A})$.

The equation for $\wh A$ is
\beq\begin{split}
&\frac1{\wh\f} = \frac{\wh A}{1-m} \frac{d\GG_m(\wh A)}{d\wh A} = \FF_m(\wh A) = \frac{m \wh A}{1-m}  \int_{-\io}^\io dy \, e^{y} 
 q(\wh A;y)^{m-1} \frac{\partial q(\wh A;y)}{\partial \wh A} 
\end{split}\eeq
At $m=1$ the equation is
\beq\label{eq:F1phi}
\frac1{\wh\f} =  \FF_1(\wh A) = -\wh A \int_{-\io}^\io dy \, e^{y} 
 \log [q(\wh A;y)] \frac{\partial q(\wh A;y)}{\partial \wh A} 
\eeq
hence the dynamical transition is reached when
$1/\wh\f_d =  \max_{\wh A} \FF_1(\wh A) $, as stated in Eq.~\eqref{phid}.
The non-trivial part of the Franz-Parisi potential, for $d\to\io$, is proportional to
\beq\label{eq:VFPdio}
\begin{split}
\VV_{\rm FP}(\wh A) &= - \log(\wh A) -  \wh\f \HH_1(\wh A) \ , \\
\HH_1(\wh A) & = \left[ m \frac{\partial \GG_m(\wh A)}{\partial m} \right]_{m=1}
=   \int_{-\io}^0 dy \, e^{y} 
\left[   q(\wh A;y) \log q(\wh A;y)  + \wh\ee  y^2 
e^{ -  \wh\ee  y^2   }  \right]  \\
&+  \int_{0}^{ \wh\s} dy \, e^{y} 
\left[   q(\wh A;y) \log q(\wh A;y)  -
\wh U_0 e^{ \wh U_0 }  \right] 
+   \int_{ \wh\s }^\io dy \, e^{y} 
\left[   q(\wh A;y) \log q(\wh A;y)   \right] 
\end{split}\eeq

\subsection{Square-well potential}

We consider here the case $\wh\ee = \io$ corresponding to a square well potential.
In this case the function $q(\wh A;y)$ is
\beq\begin{split}
q(\wh A;y) &= (1-e^{\wh U_0}) \Th\left(\frac{y+\wh A-\wh \s}{2\sqrt{\wh A}}\right)
+ e^{\wh U_0}  \Th\left(\frac{y+\wh A}{2\sqrt{\wh A}}\right)
 \ , \\
\frac{\partial q(\wh A;y)}{\partial \wh A} & = (1-e^{\wh U_0}) \frac{\wh A-y+\wh\s}{4 \wh A^{3/2} \sqrt{\pi}} e^{-\frac{(\wh A + y -\wh \s)^2}{4 \wh A}} 
+ e^{\wh U_0} \frac{\wh A-y}{4 \wh A^{3/2} \sqrt{\pi}} e^{-\frac{(\wh A + y )^2}{4 \wh A}} \ , 
\end{split}\eeq
which allow one to compute easily $\FF_1(\wh A)$ and solve the equation for $\wh A$.

\subsection{Sticky spheres}

A special case is the limit $\wh \s \to 0$ and $\wh U_0 \to \io$ with $\wh \s e^{\wh U_0} = e^{-\mu}$ or $\mu = -\wh U_0 - \log\wh \s$. In this case
\beq\begin{split}
q(\wh A;y) &=   \Th\left(\frac{y+\wh A}{2\sqrt{\wh A}}\right) + e^{-\mu} \frac{ e^{-\frac{(\wh A + y )^2}{4 \wh A}} }{\sqrt{4 \pi \wh A}}  
 \ , \\
\frac{\partial q(\wh A;y)}{\partial \wh A} & = 
 \frac{\wh A-y}{4 \wh A^{3/2} \sqrt{\pi}} e^{-\frac{(\wh A + y )^2}{4 \wh A}} 
 - e^{-\mu} \frac{2\wh A+\wh A^2 - y^2}{8 \wh A^{5/2} \sqrt{\pi}} e^{-\frac{(\wh A + y )^2}{4 \wh A}} 
 \ , 
\end{split}\eeq
In this limit the attractive glass has $\wh A = 0$, hence it is useful to compute the limit $\wh A \to 0$. We have
\beq\begin{split}
&\wh A \frac{\partial q(\wh A;y)}{\partial \wh A} \to 
 e^{-\mu} \frac{ y^2 - 2 \wh A}{4 \wh A}   e^{-y/2}  \frac{ e^{-\frac{y^2 }{4 \wh A}} }{\sqrt{4 \pi \wh A}}  \ , \\
&q(\wh A;z \sqrt{\wh A}) \to  e^{-\mu} \frac{ e^{-\frac{z^2}{4}} }{\sqrt{4 \pi \wh A}} \ ,
\end{split}\eeq
hence we can write, with $y = \sqrt{\wh A} z$: 
\beq\begin{split}
\FF_1(\wh A \to 0) & = - \int_{-\io}^\io dz \, 
 \log [q(\wh A;z \sqrt{\wh A})] e^{-\mu} \frac{ z^2 - 2}{4 }   \frac{ e^{-\frac{z^2 }{4}} }{\sqrt{4 \pi}}  \\
 & =  e^{-\mu}\int_{-\io}^\io dz \, \frac{z^2}4  \frac{ z^2 - 2}{4 }   \frac{ e^{-\frac{z^2 }{4}} }{\sqrt{4 \pi}} = \frac12 e^{-\mu}
\end{split}\eeq
The branch of the dynamical transition that corresponds to the attractive glass therefore reads
\beq\label{phid0}
\frac1{\wh\f^{\rm a}_{\rm d}} = \frac12 e^{-\mu} \ .
\eeq

\bibliographystyle{iopart-num}
\bibliography{HS}

\providecommand{\newblock}{}
\begin{thebibliography}{10}
\expandafter\ifx\csname url\endcsname\relax
  \def\url#1{{\tt #1}}\fi
\expandafter\ifx\csname urlprefix\endcsname\relax\def\urlprefix{URL }\fi
\providecommand{\eprint}[2][]{\url{#2}}

\bibitem{TS10}
Torquato S and Stillinger F~H 2010 {\em Rev. Mod. Phys.\/} {\bf 82} 2633--2672

\bibitem{crowding}
Rivas G, Ferrone F and Herzfeld J 2004 {\em EMBO Rep.\/} {\bf 5} 23--27

\bibitem{Hamming}
Hamming R~W 1997 {\em The Art of Doing Science and Engineering\/} (Gordon and
  Breach)

\bibitem{ConwaySloane}
Conway J~H and Sloane N~J~A 1993 {\em Sphere Packings, Lattices and Groups\/}
  (New York: Spriger-Verlag)

\bibitem{Co10}
Cohn H 2010 {\em Proceedings of the International Congress of Mathematicians\/}
  {\bf IV} 2416--2443 (\textit{Preprint} \eprint{{\tt arXiv:1003.3053}})

\bibitem{TJ10}
Torquato S and Jiao Y 2010 {\em Physical Review E\/} {\bf 82} 061302

\bibitem{AS12}
Andreanov A and Scardicchio A 2012 {\em Physical Review E\/} {\bf 86} 041117

\bibitem{MT13}
Marcotte E and Torquato S 2013 {\em {\tt arXiv.org:1304.5003}\/}

\bibitem{CIPZ11}
Charbonneau P, Ikeda A, Parisi G and Zamponi F 2011 {\em Phys. Rev. Lett.\/}
  {\bf 107}(18) 185702

\bibitem{VdW73}
Van~der Waals J~D 1873 {\em Doctoral thesis\/}

\bibitem{BM60}
Bernal J and Mason J 1960 {\em Nature\/} {\bf 188} 910--911

\bibitem{Sc02}
Sciortino F 2002 {\em Nature materials\/} {\bf 1} 145

\bibitem{KPZ12}
Kurchan J, Parisi G and Zamponi F 2012 {\em Journal of Statistical Mechanics:
  Theory and Experiment\/} {\bf 2012} P10012

\bibitem{noiEPL}
Sellitto M and Zamponi F 2013 {\em {\tt arXiv.org:1306.2912}, to appear on
  Europhysics Letters\/}

\bibitem{FP98}
Franz S and Parisi G 1998 {\em Physica A: Statistical Mechanics and its
  Applications\/} {\bf 261} 317--339

\bibitem{FP99}
Frisch H~L and Percus J~K 1999 {\em Phys. Rev. E\/} {\bf 60} 2942--2948

\bibitem{Fa99}
Fabbian L, G\"otze W, Sciortino F, Tartaglia P and Thiery F 1999 {\em Phys.
  Rev. E\/} {\bf 59}(2) R1347--R1350

\bibitem{ZP09}
Zaccarelli E and Poon W~C 2009 {\em Proceedings of the National Academy of
  Sciences\/} {\bf 106} 15203--15208

\bibitem{FSZT04}
Foffi G, Sciortino F, Zaccarelli E and Tartaglia P 2004 {\em Journal of
  Physics: Condensed Matter\/} {\bf 16} S3791

\bibitem{Da00}
Dawson K, Foffi G, Fuchs M, G\"otze W, Sciortino F, Sperl M, Tartaglia P,
  Voigtmann T and Zaccarelli E 2000 {\em Phys. Rev. E\/} {\bf 63}(1) 011401

\bibitem{EB02}
Eckert T and Bartsch E 2002 {\em Phys. Rev. Lett.\/} {\bf 89}(12) 125701

\bibitem{Ph02}
Pham K~N, Puertas A~M, Bergenholtz J, Egelhaaf S~U, Moussa{\i}d A, Pusey P~N,
  Schofield A~B, Cates M~E, Fuchs M and Poon W~C 2002 {\em Science\/} {\bf 296}
  104--106

\bibitem{Se10}
Sellitto M, De~Martino D, Caccioli F and Arenzon J~J 2010 {\em Phys. Rev.
  Lett.\/} {\bf 105} 265704

\bibitem{Se12}
Sellitto M 2012 {\em Phys. Rev. E\/} {\bf 86} 030502

\bibitem{KT87}
Kirkpatrick T~R and Thirumalai D 1987 {\em Phys. Rev. Lett.\/} {\bf 58}
  2091--2094

\bibitem{KW87b}
Kirkpatrick T~R and Wolynes P~G 1987 {\em Phys. Rev. B\/} {\bf 36} 8552--8564

\bibitem{WL12}
Wolynes P and Lubchenko V (eds) 2012 {\em Structural Glasses and Supercooled
  Liquids: Theory, Experiment, and Applications\/} (Wiley)

\bibitem{CCN04}
Caiazzo A, Coniglio A and Nicodemi M 2004 {\em Phys. Rev. Lett.\/} {\bf 93}(21)
  215701

\bibitem{CL06}
Crisanti A and Leuzzi L 2006 {\em Phys. Rev. B\/} {\bf 73}(1) 014412

\bibitem{Kr07}
Krakoviack V 2007 {\em Phys. Rev. B\/} {\bf 76}(13) 136401

\bibitem{CL07}
Crisanti A and Leuzzi L 2007 {\em Phys. Rev. B\/} {\bf 76}(13) 136402

\bibitem{KTZ08}
Krzakala F, Tarzia M and Zdeborov\'{a} L 2008 {\em Physical Review Letters\/}
  {\bf 101} 165702 (pages~4)

\bibitem{MP99}
M\'ezard M and Parisi G 1999 {\em The Journal of Chemical Physics\/} {\bf 111}
  1076--1095

\bibitem{FPRR11}
Franz S, Parisi G, Ricci-Tersenghi F and Rizzo T 2011 {\em The European
  Physical Journal E\/} {\bf 34} 1--17

\bibitem{VPR06}
Velenich A, Parola A and Reatto L 2006 {\em Physical Review E\/} {\bf 74}
  021410 (pages~7)

\bibitem{PZ10}
Parisi G and Zamponi F 2010 {\em Rev. Mod. Phys.\/} {\bf 82} 789--845

\bibitem{LS90}
Lubachevsky B~D and Stillinger F~H 1990 {\em J.~Stat.~Phys.\/} {\bf 60}
  561--583

\bibitem{SDST06}
Skoge M, Donev A, Stillinger F~H and Torquato S 2006 {\em Physical Review E\/}
  {\bf 74} 041127 (pages~11)

\bibitem{VCFC09}
van Meel J~A, Charbonneau B, Fortini A and Charbonneau P 2009 {\em Phys. Rev.
  E\/} {\bf 80} 061110

\bibitem{Va11}
Vance S 2011 {\em Advances in Mathematics\/} {\bf 227} 2144 -- 2156

\bibitem{Ba92}
Ball K 1992 {\em International Math. Research Notices\/} {\bf 1992} 217--221

\bibitem{KL78}
Kabatiansky G~A and Levensthein V~I 1978 {\em Problems on Information
  Transmission\/} {\bf 14} 1--17

\bibitem{PZ06a}
Parisi G and Zamponi F 2006 {\em Journal of Statistical Mechanics: Theory and
  Experiment\/} {\bf 2006} P03017

\bibitem{MS06}
Montanari A and Semerjian G 2006 {\em J.Stat.Phys.\/} {\bf 125} 23--54

\bibitem{CCPZ12}
Charbonneau P, Corwin E~I, Parisi G and Zamponi F 2012 {\em Phys. Rev. Lett.\/}
  {\bf 109}(20) 205501

\bibitem{ZK07}
Zdeborov\'{a} L and Krzakala F 2007 {\em Physical Review E\/} {\bf 76} 031131
  (pages~29)

\bibitem{KK07}
Krzakala F and Kurchan J 2007 {\em Physical Review E\/} {\bf 76} 021122
  (pages~13)

\bibitem{KZ10}
Krzakala F and Zdeborov{\'a} L 2010 {\em EPL (Europhysics Letters)\/} {\bf 90}
  66002

\bibitem{KW87}
Kirkpatrick T~R and Wolynes P~G 1987 {\em Phys. Rev. A\/} {\bf 35} 3072--3080

\bibitem{KT88}
Kirkpatrick T~R and Thirumalai D 1988 {\em Phys. Rev. A\/} {\bf 37}(11)
  4439--4448

\bibitem{KT89}
Kirkpatrick T~R and Thirumalai D 1989 {\em Journal of Physics A: Mathematical
  and General\/} {\bf 22} L149

\bibitem{FP95}
Franz S and Parisi G 1995 {\em Journal de Physique I\/} {\bf 5} 1401--1415

\bibitem{FP97}
Franz S and Parisi G 1997 {\em Phys. Rev. Lett.\/} {\bf 79} 2486--2489

\bibitem{CFP98}
Cardenas M, Franz S and Parisi G 1998 {\em Journal of Physics A: Mathematical
  and General\/} {\bf 31} L163--L169

\bibitem{Mo95}
Monasson R 1995 {\em Phys. Rev. Lett.\/} {\bf 75} 2847--2850

\bibitem{DawFraSel03}
Dawson K~A, Franz S and Sellitto M 2003 {\em Europhysics Letters\/} {\bf 64}
  302--308

\bibitem{CCGGGPV10}
Cammarota C, Cavagna A, Giardina I, Gradenigo G, Grigera T~S, Parisi G and
  Verrocchio P 2010 {\em Phys. Rev. Lett.\/} {\bf 105}(5) 055703

\bibitem{MP96}
M{\'e}zard M and Parisi G 1996 {\em Journal of Physics A: Mathematical and
  General\/} {\bf 29} 6515

\bibitem{MP09}
Mezard M and Parisi G 2012 {\em Structural Glasses and Supercooled Liquids:
  Theory, Experiment and Applications\/} ed PGWolynes and VLubchenko (Wiley \&
  Sons) (\textit{Preprint} \eprint{{\tt arXiv:0910.2838}})

\bibitem{BJZ11}
Berthier L, Jacquin H and Zamponi F 2011 {\em Phys. Rev. E\/} {\bf 84}(5)
  051103

\bibitem{BF99}
Bergenholtz J and Fuchs M 1999 {\em Phys. Rev. E\/} {\bf 59}(5) 5706--5715

\bibitem{Go09}
G{\"o}tze W 2009 {\em Complex dynamics of glass-forming liquids: A
  mode-coupling theory\/} (Oxford University Press, USA)

\bibitem{STZ03}
Sciortino F, Tartaglia P and Zaccarelli E 2003 {\em Phys. Rev. Lett.\/} {\bf
  91}(26) 268301

\bibitem{CR07}
Charbonneau P and Reichman D~R 2007 {\em Phys. Rev. Lett.\/} {\bf 99}(13)
  135701

\bibitem{CK93}
Cugliandolo L~F and Kurchan J 1993 {\em Phys. Rev. Lett.\/} {\bf 71} 173--176

\bibitem{CC05}
Castellani T and Cavagna A 2005 {\em Journal of Statistical Mechanics: Theory
  and Experiment\/} {\bf 2005} P05012

\bibitem{IM10}
Ikeda A and Miyazaki K 2010 {\em Phys. Rev. Lett.\/} {\bf 104} 255704

\bibitem{SS10}
Schmid B and Schilling R 2010 {\em Phys. Rev. E\/} {\bf 81} 041502

\bibitem{Sz10}
Szamel G 2010 {\em Europhysics Letters\/} {\bf 91} 56004

\bibitem{CIPZ12}
Charbonneau P, Ikeda A, Parisi G and Zamponi F 2012 {\em Proceedings of the
  National Academy of Sciences\/} {\bf 109} 13939--13943

\bibitem{Hansen}
Hansen J~P and McDonald I~R 1986 {\em Theory of simple liquids\/} (London:
  Academic Press)

\end{thebibliography}

\end{document}